\documentclass[prl,twocolumn,dvipdf,amsmath,amssymb,floatfix]{revtex4}
\usepackage{bm}
\usepackage{graphicx}

\begin{document}

\newcommand{\fmov}{f_\mathrm{mov}}
\newcommand{\gamman}{\gamma_0}
\newcommand{\ngrad}{N_\mathrm{grad}}

\title{Generic transient memory formation in disordered systems with noise}

\author{Nathan~C.~Keim}
\altaffiliation[Present address: ]{Department of Mechanical Engineering and Applied Mechanics, University of Pennsylvania, 220 S.\ 33rd St., Philadelphia, PA 19104}
\affiliation{James Franck Institute and Department of Physics, University of Chicago,
929 E.\ 57th St., Chicago, IL 60637, USA}
\author{Sidney~R.~Nagel}
\email{srnagel@uchicago.edu}
\affiliation{James Franck Institute and Department of Physics, University of Chicago,
929 E.\ 57th St., Chicago, IL 60637, USA}

\date{\today}

\begin{abstract}

Out-of-equilibrium disordered systems may form memories of external driving in a remarkable fashion. The system ``remembers'' multiple values from a series of training inputs yet ``forgets'' nearly all of them at long times despite the inputs being continually repeated. Here, learning and forgetting are inseparable aspects of a single process. The memory loss may be prevented by the addition of noise. We identify a class of systems with this behavior, giving as an example a model of non-brownian suspensions under cyclic shear.

\end{abstract}

\maketitle

Systems render information about their formation inaccessible to observers after they relax to equilibrium; a system that has not fully relaxed has the potential to retain memories of its creation. Such behavior raises questions about the type and amount of information preserved, as well as the basic operations of memory: imprinting, reading and erasure of information. Here we describe a class of systems that combine storage, reading, and loss in a single, uniform process. In the short term, these systems form concurrent memories of multiple external driving parameters. However, with no change in the driving, the systems gradually eliminate this information, selecting only one or two input values to be preserved at long times. With the addition of noise, all memories can be retained indefinitely. 

Such surprising behavior had first been found in a model of electronic transport by sliding charge-density waves~\cite{coppersmith97,povinelli99}. However, it was not clear whether this memory formation is unique to that system or if it is an example of a more generic phenomenon. When a charge-density wave is driven across a sample by a series of discrete voltage pulses, each of the same duration $A$, the current response eventually becomes phase-locked to the end of each pulse~\cite{fleming86,brown86}. The response therefore reveals information about the training history. This ``pulse-duration memory'' depends only on the driving with no fine-tuning of parameters. This behavior was modeled as self-organization of the charge-density wave around random defects in the material~\cite{coppersmith87}. Further work modeled the behavior of the system when $M$ pulse durations ($A_1$,$A_2$ \dots $A_M$) were applied in an arbitrary, repeating pattern and showed that the system learns all these inputs at intermediate times~\cite{coppersmith97,povinelli99}. However, as the learning progresses, most of the responses diminish, until eventually only two memories remain. If noise is added, all the memories persist indefinitely. 

Since that work, this memory formation has remained unique to charge-density waves; despite the ubiquity of cyclically driven disordered systems, no one has addressed if multiple transient memory formation could be generic. Here we show that it is. A commonly observed phenomenon, not previously interpreted in terms of memory formation, has similar behavior to the multiple-pulse-duration memory in charge-density waves. Our finding thus points to a new class of memory in disordered systems.

When a disordered system, \emph{e.g.} foam, granular material, or suspension, undergoes oscillatory shear, the individual particles rearrange as they are forced to traverse energy barriers into nearby wells. When returned to the initial, zero-strain position, the system has reorganized. If this is done repeatedly at a fixed strain amplitude, $\gamma_0$, the system anneals during a transient relaxation period, followed by a steady state where further applications of $\gamma_0$ produce no rearrangements. This is a common occurrence, seen, for example, in the density and crystallization of granular media~\cite{toiya04,mueggenburg05}, in ordering of colloids~\cite{ackerson88,haw98pre}, in particle diffusion in viscous suspensions~\cite{pine05,corte08,menon09}, or in plastic events in amorphous materials~\cite{lundberg08}. As a steady state is approached, fewer particles rearrange per cycle. As long as the system is irreversible, it explores a different configuration with each cycle; it keeps exploring possible states until it finds, if possible, a reversible one which will then be the steady state. In general, larger-amplitude strains cause longer relaxation times to the steady state~\cite{haw98pre,toiya04,corte08}. 

We can think of the steady-state response as a memory. After many training cycles with amplitude $\gamma_i$, the system may reach a steady state where changes no longer occur. If the system has found a reversible state for $\gamma_i$, it must also be stable for $\gamma_j$ if $\gamma_j < \gamma_i$, since a larger-amplitude shear encompasses smaller ones on its route.  Thus, once in a steady-state configuration for $\gamma_i$, a smaller amplitude $\gamma_j$ does not alter that state.  However, if the system is in a steady state for $\gamma_j$, application of $\gamma_i$ will eventually erase that steady-state configuration. Thus, the steady-state configuration is a memory of a specific strain amplitude. It can be read out simply by observing at what strain the system starts to be irreversible. 

\begin{figure}
\begin{center}
\includegraphics[width=3.3in]{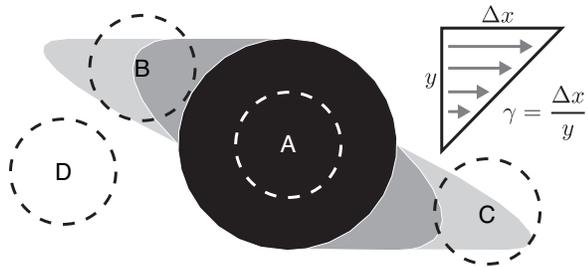}
\end{center}
\caption{Simulation algorithm after~\cite{corte08}. Particle $A$, outlined in dashes, and its neighbors ($B$, $C$, $D$) are sheared to $+\gamma$ and then returned to $\gamma=0$ (defined in diagram). The center of each particle lies in a shaded region corresponding to a strain that would bring it into collision with $A$: $\gamma=0$ (black), 1 (dark grey), and 2 (light grey). At $\gamma = 1$, $A$ and $B$ collide; at $\gamma=2$, $C$ also collides. $D$ never collides with $A$ for any $\gamma > 0$. After all particles are considered in this fashion, colliding particles are given small random displacements and the algorithm repeats.}
\label{fig_algorithm}
\end{figure}

To demonstrate this behavior, we use a model developed by Cort\'e \emph{et al.}\ to model viscous, non-brownian suspensions in the limit of zero inertia under cyclic shear~\cite{corte08}. That model simulates the rearrangement of particles as they pass close to each other during a shearing cycle of amplitude $\gamma_0$. It evolves the positions of $N$ discs with diameter $d$ in a 2-dimensional box of area $A_\mathrm{box}$ with periodic boundary conditions. The algorithm considers the effect on each particle of applying a uniform strain $\gamma$ as illustrated in Fig.~\ref{fig_algorithm}, simply translating the center of each particle by $\Delta x = y \gamma$. If a particle overlaps with another at any strain $0 \leq \gamma \leq \gamma_0$ along the motion, it is tagged. To advance to the next timestep, each tagged particle is moved in a random direction, by a distance with a uniform random distribution between 0 and $\epsilon d$. This model does a remarkable job at reproducing the phenomena observed in the experiment~\cite{pine05,corte08}: for sufficiently small strain, repeated application of $\gamma_0$ eventually causes the system to become reversible so that further application of $\gamma_0$ no longer changes particle positions. Moreover, in both experiment and the model there is a critical strain amplitude $\gamma_c$, above which complete self-organization does not occur. 

In our simulations, $N=10^4$, the box is square, and the area fraction $\phi \equiv (N \pi d^2 /4) /A_\mathrm{box}= 0.2$ so that $\gamma_c \approx 4$. We focus on systems that reach a steady state: $\gamma < \gamma_c$. We use much smaller random displacements than Cort\'e \emph{et al.}\ to evolve the system ($0.005 \leq \epsilon \leq 0.1$ \emph{vs.}\ 0.5), requiring simulation runs of $> 10^6$ cycles. In the limit $\epsilon \ll 1$, we find that the evolution time scales as $\epsilon^{-2}$. 

\begin{figure}
\begin{center}
\includegraphics[width=3.3in]{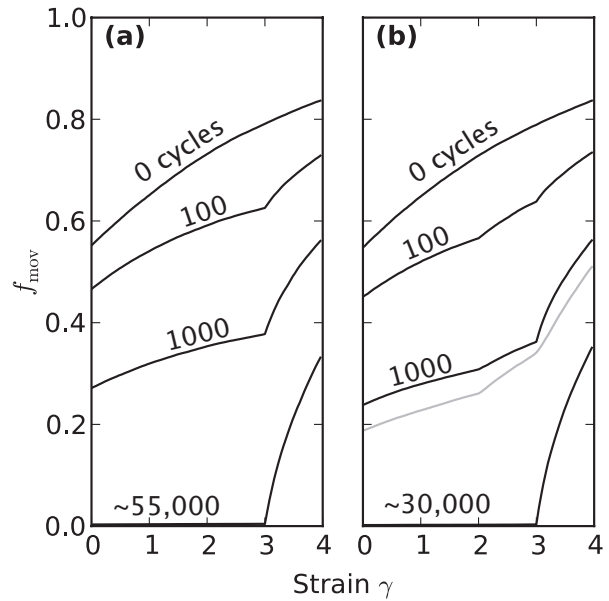}
\end{center}
\caption{Fraction of particles moved $\fmov$ versus trial strain $\gamma$, at selected times during the system's evolution. \textbf{(a)}~Evolution with a single training amplitude, $\gamma_1 = 3.0$. After 100 and 1000 cycles, the system's self-organization is incomplete, but the value of $\gamma_1$ can be readily identified from kinks in each curve. \textbf{(b)}~With dual training values, $\gamma_1 = 3.0$ and $\gamma_2 = 2.0$, (pattern: $\gamma_1, \gamma_2,\gamma_2, \gamma_2, \gamma_2, \gamma_2$, repeat \ldots), both values can be identified at intermediate times. The system completely self-organizes, retaining only the larger training value, $\gamma_1 = 3.0$, after $\sim$ 30,000 cycles. Grey line: memory of both values remains after $10^5$ cycles when the system is stabilized by noise ($\epsilon_\mathrm{noise} = 0.006$). Plots are averaged over 9 runs of $N = 10^4$, with $\epsilon = 0.1$.}
\label{fig_training}
\end{figure}

In Fig.~\ref{fig_training}, we plot the fraction of particles $\fmov$ that would be moved by the algorithm in a single cycle, versus the strain. Because a particle is moved only when the shearing motion brings it into contact with other particles, each curve probes the separations between particles, averaged over the entire system. Fig.~\ref{fig_training}a shows the evolution of $\fmov(\gamma)$ as the system gradually self-organizes from a random configuration, under a single applied strain amplitude $\gamma_1=3$. This value is significantly less than the critical strain for this packing fraction, $\gamma_c \approx 4$, so that a steady state can be formed. Crucially, when the training process of repeatedly shearing by $\gamma_1$ is complete, a shear with any $\gamma \leq \gamma_1$ results in no rearrangement of the particles. This permits the memory to be read out, without knowledge of the system's preparation, by applying a cyclic shear with progressively larger trial $\gamma$ until rearrangement is observed. However, even before self-organization is complete, the memory may be read out by observing a marked increase in the irreversibility of the system quantified by the change in slope of $\fmov$ at $\gamma_1$. 

Progressing to two simultaneous memories, we encounter a crucial distinguishing question in evaluating memory in disordered systems: can a memory generally be added without erasing another? Figure~\ref{fig_training}b shows that the same system can be trained with 2 memories at once, combined in a repeating pattern of $\gamma_1=3$ and $\gamma_2=2$. (We repeat the smaller amplitude, $\gamma_2=2$ five times for every application of $\gamma_1=3$.  This helps make the memory of $\gamma_2=2$ more apparent. The memory would be there, only harder to see, if we used equal numbers of $\gamma_1$ and $\gamma_2$ in each cycle.) This result distinguishes this type of memory from (i) ``return-point memory'' in magnets~\cite{barker83,sethna93} where application of a large magnetic field immediately erases all stored memories of smaller fields, or from (ii) aging, rejuvenation and memory in glasses~\cite{jonason98,zou10} where the annealing protocol must be done at progressively lowered temperatures in order to be remembered upon reheating. In contrast, in our system, when a cycle with $\gamma_i$ is followed by one with any larger or smaller $\gamma_j$, we find that the incipient memory of $\gamma_i$ is only partly disrupted: Fig.~\ref{fig_training}b shows that both $\gamma_1$ and $\gamma_2$ are evident in $\fmov(\gamma)$ for much of the system's evolution. We emphasize that all memories are observed at all times of the transient period, regardless of which value was most recently applied. Using smaller values of $\epsilon$ improves multiple-memory formation. 

\begin{figure}
\begin{center}
\includegraphics[width=3.5in]{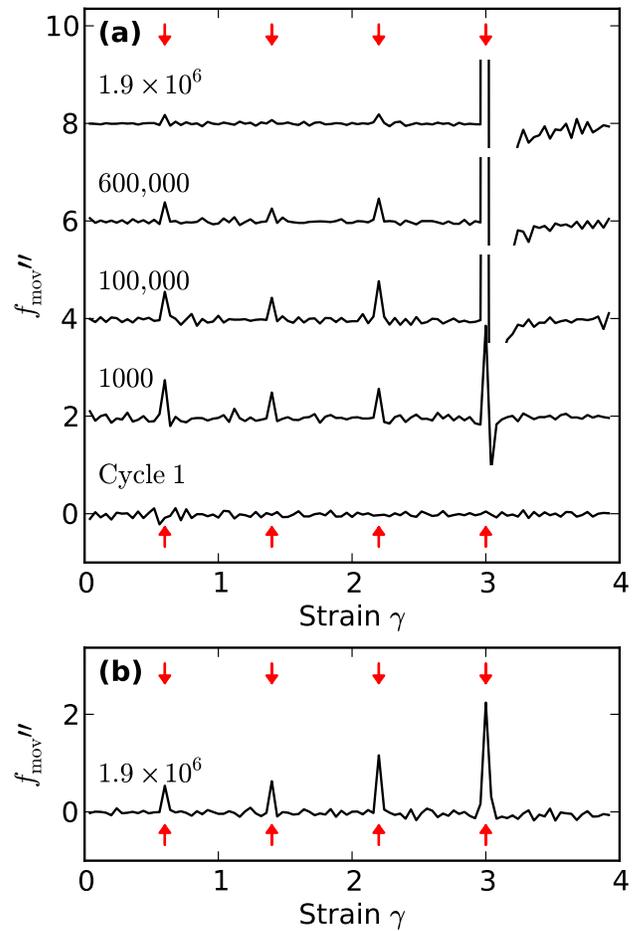}
\end{center}
\caption{
\textbf{(a)} Snapshots of memory strength $\fmov'' \equiv d^2 \fmov / d\gamma^2$ versus $\gamma$ for a training process of 4 shear values, shown by the arrows at the top and bottom. The training pattern is 3.0, 0.6, 0.6, 2.2, 2.2, 0.6, 1.4, 1.4, 0.6. Each curve is displaced vertically by 2 above the previous one. The simulations used $\epsilon = 0.005$. The memories are not present initially (labeled Cycle 1) when the training is begun from a random configuration; by cycle 1000 memories appear at all 4 positions; at 600,000 cycles the memories at the 3 lower training values are weaker and after $1.9 \times 10^6$ cycles they are indistinct. In the curves at $10^5$, $6 \times10^5$ and $1.9 \times 10^6$ cycles, the peak at $\gamma = 3.0$ has grown too large to plot on this graph; its actual height is 23.4, 52.4, and 73.2, respectively. \textbf{(b)} Adding small noise preserves memory indefinitely; here we evolved the system through $1.9 \times 10^6$ cycles with $\epsilon_\mathrm{noise} = 4 \times 10^{-4}$. To compute the curves, $\fmov$ was sampled at intervals of $\Delta \gamma = 0.04$.}
\label{fig_peaks}
\end{figure}

To quantify the evolution and strength of each memory, we compute the second derivative of curves like those in Fig.~\ref{fig_training}, \emph{i.e.}\ $\fmov'' \equiv d^2 \fmov / d\gamma^2$. A peak in $\fmov''(\gamma)$ signifies a memory because the rapid increase in the slope of $\fmov$ as the system is strained past the training value is a sign of increasing irreversibility. Fig.~\ref{fig_peaks}a shows the evolution of 4 independent memories. We use a system that evolves slowly, $\epsilon = 0.005$, and average $\fmov''$ over 117 runs with $N = 10^4$.

In all our simulations, the memories are all present throughout the transient period and, as the system evolves, memory of all but the highest training value is eventually lost. In Fig.~\ref{fig_training}, the smaller $\gamma_2=2$ is banished when the system completely self-organizes after $\sim 30,000$ cycles. Fig.~\ref{fig_peaks}a shows that for the case of 4 memories, the memories of the three smallest $\gamma_i$ become nearly indistinct after $\sim 10^6$ cycles. 

In order for multiple transient memories to be stored and retrieved in our system, the information must be stored locally. We can see how this works by using a one-dimensional version of the algorithm~\cite{corte08}. Here the memories are preserved in the spacing between particles, the local density.  Inside a region of the disordered sample that is initially more dense than the average, small particle rearrangements leave the local density unchanged. The local density can change only when the entire region has expanded. A memory of a small value, $\gamma_i$, can be retained in this region transiently. In order for that memory to be erased, the high-density region must dissolve as its boundaries shrink.  As this is a slow process, memories can persist transiently even though a more stable (\emph{i.e.}, larger strain amplitude) memory is growing elsewhere in the system. This is similar to the multiple pulse-duration memories in charge-density waves~\cite{coppersmith97,povinelli99}.

As with the charge-density wave model~\cite{coppersmith97,povinelli99}, we find that adding noise to the system can stabilize multiple memories indefinitely. In our present model, we show this by applying a random kick to each particle at each time step, regardless of whether it has collided with another particle. The kick is drawn from a two-dimensional Gaussian distribution with standard deviation $\epsilon_\mathrm{noise}d$; we find memories are best stabilized by $\epsilon_\mathrm{noise} \sim 0.1 \epsilon$. The preservation of memories by noise is shown by the grey curve in Fig.~\ref{fig_training}b, and by Fig.~\ref{fig_peaks}b. Although at first this may seem surprising, we can easily understand this behavior because noise disrupts the self-organization and thus ensures that the system never reaches its final fixed point. The presence of noise thus insures that the system is constrained never to leave the transient regime. 

The simulation algorithm we have described was originally developed to explain the behavior of a cyclically sheared non-brownian suspension~\cite{corte08}. This suggests that this experimental system may exhibit the kind of memory we have demonstrated in simulations. The memories could be formed and then read out with a strain-controlled, strain-resolved measurement in a rheometer. However, the experiments of Cort\'e \emph{et al.}~\cite{corte08} have a characteristic self-organization time $\tau_0 \alt 50$ cycles, so that a detection approaching the quality of that presented here may be difficult in that realization. 

We can contrast this form of memory with some others that have been proposed in materials and biology. Return-point memory in magnets also has a hierarchy of training inputs. But those systems differ in that they organize promptly and any memory is wiped out as soon as a larger field is applied~\cite{barker83,sethna93}. This strict ordering of memories is also true for aging and rejuvenation in glasses in that a higher temperature wipes out memories at lower temperatures~\cite{jonason98,zou10}. This differs from multiple transient memories. The effect of noise in enhancing memory retention, rather than degrading it, is also a remarkable result of the memory mechanism presented here and has no counterpart of which we are aware in those other systems. Lastly, multiple transient memories can be contrasted with those of the neural-networks used to model associative memory~\cite{hopfield82}. In the former case, memory formation is local and depends on the path taken during each cycle. If a section of the system is removed from its surroundings, it would continue to have the memories stored within it, although with perhaps somewhat degraded resolution. 

Under repeating driving, our simulation of sheared particles stores multiple memories, but eventually retains only one. This also represents a simple information processing operation, choosing and storing the largest value among a repeated set of inputs. Our system shares this response with charge-density waves~\cite{coppersmith97,povinelli99}. However, it shares only a few attributes of the underlying dynamics: unlike the charge-density wave model, it is not deterministic and is not confined to one dimension. The present findings suggest that this behavior represents a new class, which instead of being limited to charge-density-wave conductors, may be found in disordered systems generally --- including possibly biological systems where cyclic behavior is common and noise is important. In particular, there are many annealing protocols for disordered materials; as mentioned above, under oscillatory shear, or vibration~\cite{knight95}, many disordered materials relax towards a steady state. One of the outcomes of this study is to suggest that these generic steady states can be considered as memories of the annealing process and that systems such as these are complex enough to store several memories simultaneously.  

We thank Susan Coppersmith, Michelle Povinelli and Wendy Zhang for many inspiring discussions. This work was supported by NSF Grant DMR-1105145. Use of computation facilities of the NSF-MRSEC and of the DOE Office of Basic Energy Sciences DE-FG02-03ER46088 are gratefully acknowledged.

\bibliography{references-nourl}

\end{document}